\documentclass[twocolumn,showpacs]{revtex4-1}

\usepackage{newlfont}
\usepackage[below]{placeins}
\usepackage{flafter}
\usepackage{amsfonts}
\usepackage{amsmath}
\usepackage{amssymb}
\usepackage[american]{babel}
\usepackage{graphicx}
\usepackage{subfigure}
\usepackage{multirow}
\usepackage{url}
\usepackage{color}
\usepackage{graphicx}
\usepackage{mathtools}
\usepackage{float}

\allowdisplaybreaks

\newcommand{\ket}[1]{\big| #1 \big\rangle}
\newcommand{\bra}[1]{\big\langle #1 \big|}
\newcommand{\braket}[2]{\big\langle #1 \big| #2 \big\rangle}                 
\newcommand{\bracket}[3]{\big\langle #1 \big| #2 \big| #3 \big\rangle}       

\begin{document}


\title{Quantum search on the two-dimensional lattice using \\
the staggered model with Hamiltonians}
\author{R. Portugal$^{1,2}$ and T. D. Fernandes$^{1,3}$}
\affiliation{ 
$^1$National Laboratory of Scientific Computing (LNCC), Petr\'{o}polis, RJ,  25651-075, Brazil\\
$^2$Universidade Cat\'olica de Petr\' opolis, Petr\'{o}polis, RJ, 25685-070,  Brazil\\
$^3$Universidade Federal do Esp\'{i}rito Santo - UFES, 29500-000, Alegre, Brazil\\
}

\date{\today}

\begin{abstract}
Quantum search on the two-dimensional lattice with one marked vertex and cyclic boundary conditions is an important problem in the context of quantum algorithms with an interesting unfolding. It avails to test the ability of quantum walk models to provide efficient algorithms from the theoretical side and means to implement quantum walks in laboratories from the practical side. In this paper, we rigorously prove that the recent-proposed staggered quantum walk model provides an efficient quantum search on the two-dimensional lattice, if the reflection operators associated with the graph tessellations are used as Hamiltonians, which is an important theoretical result for validating the staggered model with Hamiltonians. Numerical results show that on the two-dimensional lattice staggered models without Hamiltonians are not as efficient as the one described in this paper and are, in fact, as slow as classical random-walk-based algorithms.
\end{abstract}

\maketitle

\section{Introduction}

Quantum search was introduced by Grover's seminal work~\cite{Grover:1997a}, which described an evolution operator $U$ that can be written as a product of two operators $G\cdot R_0$, where $G$ is the well-known Grover operator and $R_0$ is the operator that marks one vector of the computational basis by changing its sign, for instance, if the marked element is 0 then $R_0\ket{0}=-\ket{0}$ and  $R_0\ket{i}=\ket{i}$, if $i\neq 0$. Originally, Grover presented his algorithm having database searching in mind. Soon it became evident that the algorithm has a broader scope, can be used for searching more than one element, and is the simplest example of the technique called amplitude amplification~\cite{BBHT98}. It was also realized that the Grover algorithm can be formulated as a coined quantum walk search on the complete graph~\cite{Ambainis:2005} and, recently, it was shown that Grover's algorithm is a staggered quantum walk on the complete graph using two tessellations: the first one has one polygon with all vertices and the second one has one polygon with the marked vertex only~\cite{PSFG15}.

A natural way to generalize Grover's algorithm is by analyzing the quantum search on graphs different from the complete graph. For instance, results for the two-dimensional lattice with cyclic boundary conditions were presented in Refs.~\cite{Ben02,Aaronson:2003} and for the hypercube using the coined quantum walk in Ref.~\cite{Shenvi:2003}.

Quantum search on the two-dimensional lattice using quantum walks has an interesting unfolding. Ambainis~\textit{et al.}~\cite{Ambainis:2005} used a quantum-walk-based search algorithm that finds the marked vertex in $O(\sqrt{N}\ln N)$ time after employing the method of amplitude amplification, where $N$ is the number of vertices. By adding an extra qubit to the system, Tulsi~\cite{Tul08} was able to improve the time complexity  to $O(\sqrt{N\ln N})$ without using the amplitude-amplification method. Afterwards, Ambainis~\textit{et al.}~\cite{Ambainis:2012} also showed how to eliminate amplitude amplification using the original algorithm and performing a classical post-processing search in order to obtain the time complexity $O(\sqrt{N\ln N})$ without an extra qubit. This can be considered the best quantum search on the two-dimensional lattice up to now. In this work, we present a new search algorithm on the two-dimensional lattice with the same time complexity $O(\sqrt{N\ln N})$ without using coins. An open problem is to find an algorithm with time complexity $O\big(\sqrt{N}\big)$, which would go beyond the square root of the hitting time of a random walk on the two-dimensional lattice.

Coinless quantum walks were analyzed in Refs~\cite{Patel05,Falk:2013,PBF15,Ambainis:2013} and motivated the development a new model called staggered quantum walk~\cite{PSFG15,Por16b}. The staggered model with two tessellations can exactly reproduce the evolution of all instances of Szegedy's model~\cite{Szegedy:2004} and the instances of the coined model that use the Grover or Hadamard coin. The extension with Hamiltonians was proposed in Ref.~\cite{POM16} adding more flexibility to the model. The results in the present paper could not be found without this extension. 

The evolution operator of the staggered model is the product of local operators, each one obtained from a graph tessellation~\cite{Por16b}. A tessellation $\mathcal{T}$ of a graph $\Gamma$ is a partition of the vertex set of $\Gamma$ so that each partition element (called polygon) is a clique. A clique is a subset of vertices of a graph such that any two vertices of this subset are adjacent (complete subgraph).  In order to obtain the evolution operator of a staggered quantum walk on $\Gamma$, we need to define a tessellation cover, which is a set of tessellations ${\mathcal{T}}_1,...,{\mathcal{T}}_k$ so that the union $\cup_{i=1}^k\,{\mathcal{E}}({\mathcal{T}}_i)$ is the edge set of $\Gamma$, where ${\mathcal{E}}({\mathcal{T}}_i)$ is the set of edges in tessellation ${\mathcal{T}}_i$ and $k$ is the size of the tessellation cover. Each tessellation ${\mathcal{T}}_i$ is associated with a unitary and Hermitian operator $H_i$~\cite{Por16b}; the product of operators $H_1,..., H_k$ defines the evolution operator of the staggered model; the product of operators $\exp(i\theta_1 H_1),..., \exp(i\theta_k H_k)$, where $\theta_1, ...,\theta_k$ are angles, defines the evolution operator of the staggered model with Hamiltonians~\cite{POM16}. The order of the local operators matters; a tessellation cover can be associated with more than one evolution operator.


The smallest tessellation cover of a two-dimensional square lattice with cyclic boundary conditions and width $2n$ (height $2n$), where $n$ is an integer, has size 4. Now we describe four tessellations, whose union covers the lattice, and each tessellation is composed of polygons of two vertices. Consider the set of vertices $(x,y)$, where $x$ and $y$ are labels such that $x+y$ is even and $0\le x,y<2n$ and the arithmetic is performed modulo $2n$. The first tessellation ${\mathcal{T}}_{00}$ comprises the polygons $\{(x,y),(x+1,y)\}$, that is, ${\mathcal{T}}_{00}=\{\{(x,y),(x+1,y)\}:x+y\textrm{ is even}\}$. Likewise, we define ${\mathcal{T}}_{01}=\{\{(x,y),(x,y+1)\}:x+y\textrm{ is even}\}$, ${\mathcal{T}}_{10}=\{\{(x,y),(x-1,y)\}:x+y\textrm{ is even}\}$, and ${\mathcal{T}}_{11}=\{\{(x,y),(x,y-1)\}:x+y\textrm{ is even}\}$, where $0\le x,y<2n$. Notice that each tessellation covers all vertices, are composed of cliques, and ${\mathcal{E}}({\mathcal{T}}_{00})\cup{\mathcal{E}}({\mathcal{T}}_{01})\cup{\mathcal{E}}({\mathcal{T}}_{10})\cup{\mathcal{E}}({\mathcal{T}}_{11})$ is the lattice edge set, establishing, therefore, a well-defined tessellation cover.

Each tessellation ${\mathcal{T}}_{ab}$ is associated with a unitary operator $U_{ab} = \textrm{e}^{i\,\theta H_{ab}}$, where $ab$ is either 00 or 01 or 10, or 11, $\theta$ is an angle, and $H_{ab}=2\Pi_{ab}-I$ is a reflection operator (Hermitian and unitary), where $\Pi_{ab}$ is the orthogonal projector on the subspace spanned by the vectors associated with the polygons of tessellation ${\mathcal{T}}_{ab}$~\cite{POM16}. In this work, we focus on the quantum walk whose evolution operator is the product $U = -U_{11} U_{10} U_{01} U_{00}$ with $\theta=\pi/4$. The minus sign was introduced to help the algorithm analysis. By changing the order of tessellations ${\mathcal{T}}_{01}$ and ${\mathcal{T}}_{10}$, we obtain another independent evolution operator, which is worse for searching algorithms as our numerical analysis has shown.

We can split the vertices $(x,y)$ of the two-dimensional lattice in two classes using the parity of the index sum $x+y$. If $x+y$ is even, vertex $(x,y)$ is in the first class, otherwise it is in the second class. Using those classes and exploring the translational symmetries of the two-dimensional lattice, we can define two sets of vectors $\ket{{\psi}_{k l}^{\textrm{0}}}$ and $\ket{{\psi}_{k l}^{\textrm{1}}}$ so that, for fixed $k$ and $l$ in the range $0\le k,l < N$, where $N$ is the number of vertices, their linear combination is invariant under the action of the evolution operator. This fact allows us to formulate a technique to find the spectrum of the evolution operator.

To search a marked vertex, we use the paradigm made explicit by the Grover algorithm~\cite{Grover:1997a}. The vertices are marked by a unitary operator called oracle that inverts the sign of the marked vertices. Without loss of generality (due to the translational symmetries of the two-dimensional lattice), we consider vertex $(0,0)$ as the target. This reduces the amount of calculation to analyze the searching algorithm. In this case, the oracle is given by $R_0=I-2\ket{0,0}\bra{0,0}$ and the modified evolution operator by ${\cal U}_0=U\,R_0$. In this work we prove that the time complexity for finding the marked vertex using ${\cal U}_0$ is $O(\sqrt{N\ln N})$.

Before starting to address analytically the evolution of this quantum walk, we had numerically analyzed the time complexity of many a kind of staggered quantum walks with Hamiltonians on the two-dimensional lattice using four tessellations~\cite{FP16}. The main conclusions were that the original staggered model (with $\theta=\pi/2$) has no instance that finds the marked vertex quicker than random-walk-based algorithms even taking non-uniform vectors associated with the polygons. After trying many values of $\theta$, the numerical results pointed out that two models had improvement over classical algorithms: the first is the one analytically addressed in this paper, which finds the marked vertex in $O(\sqrt{N\ln N})$ time with $\theta=\pi/4$, and the second is the one using the evolution operator  $-U_{11} U_{01} U_{10} U_{00}$ (the order is permuted) also with $\theta=\pi/4$, which has time complexity $\Theta(N^{3/4})$, established via numerical methods. In both cases, the time complexity quickly deteriorates when the value of $\theta$ moves away from $\pi/4$.

The structure of the paper is as follows. In Sec.~\ref{sec1}, we describe the algorithm that efficiently finds one marked vertex in a two-dimensional lattice using the staggered model with Hamiltonians. In Sec.~\ref{sec2}, we find the spectrum of the evolution operator using the Fourier analysis when there is no marked vertex. In Sec.~\ref{sec3}, we analyze the time complexity of the algorithm by calculating the number of steps and the success probability. In Sec.~\ref{conc}, we draw our conclusions.

\section{The Algorithm}\label{sec1}

Consider a two-dimensional lattice with $N$ vertices and cyclic boundary conditions and assume that $N=4n^2$ for some integer $n>1$. The Hilbert space associated with this lattice is ${\cal H}^N$, whose computational basis is $\{\ket{x,y}:0\le x,y <2n\}$.
 
The evolution operator based on the staggered quantum walk model with Hamiltonians~\cite{POM16} is $U = -U_{11} U_{10} U_{01} U_{00},$
where $U_{ab} = \textrm{e}^{\frac{\pi i}{4}H_{ab}}$,
\begin{equation}\label{eq:H}
	H_{ab}\,=\,2\sum_{\mathclap{\substack{x,y=0\\x+y\textrm{ even}}}}^{2n-1}\ket{u_{xy}^{(ab)}}\bra{u_{xy}^{(ab)}}-I,
\end{equation}
and
\begin{equation}\label{eq:u}
	\ket{u_{xy}^{(ab)}}\,=\,\frac{\ket{x,y}+\ket{x+(-1)^a\delta_{b0},y+(-1)^a\delta_{b1}}}{\sqrt 2}.
\end{equation}
The arithmetic inside the kets is performed modulo $(2n)$.

Without lost of generality, let us consider vertex $(x,y)=(0,0)$ as the target, which is marked by operator $R_{0}=I-2\ket{0,0}\bra{0,0}$.
The searching operator for the two-dimensional lattice (called modified evolution operator) is 
\begin{equation}
	{\cal U}_0=U R_{0}.
\end{equation}
The initial condition is
\begin{equation}\label{IC}
	\ket{\psi_0}\,=\,\frac{1}{\sqrt{2}\,n}\sum_{x+y\textrm{ even}} \ket{x,y},
\end{equation}
where the indices of the sum run on the same values of Eq.~(\ref{eq:H}). The state at time $t$ is $\ket{\psi(t)}=({\cal U}_0)^t\ket{\psi_0}$ and the probability distribution is $p_{xy}(t)=|\bra{x,y}({\cal U}_0)^t\ket{\psi_0}|^2$. 

In the next sections, we show that if the running time is $\Theta(\sqrt{N \ln N})$, the marked site will be found with probability $\Theta(1/\ln N)$.

\section{Fourier analysis}\label{sec2}

In order to analyze the performance of the algorithm, we need to calculate the eigenvalue  of ${\cal U}_0$ with the smallest positive argument and its associated eigenvector~\cite{Ambainis:2005,Tul12,Portugal:Book}. To accomplish this task, we need to find an eigenbasis of $U$ and the corresponding eigenvalues. The Fourier analysis helps in the second task. Define vectors 
\begin{eqnarray}
  \ket{\psi^{0}_{kl}}&=&\frac{1}{\sqrt{2}\,n}\sum_{x,y=0}^{n-1}\left( {\omega}^{2xk+2yl}\ket{2x,2y} +\right. \nonumber \\
	&&\left.{\omega}^{(2x+1)k+(2y+1)l}\ket{2x+1,2y+1}\right)   \label{psi_0} \\
  \ket{\psi^{1}_{kl}}&=&\frac{1}{\sqrt{2}\,n}\sum_{x,y=0}^{n-1}\left({\omega}^{2xk+(2y+1)l}\ket{2x,2y+1}  +\right.  \nonumber\\
	&&\left.{\omega}^{(2x+1)k+2yl}\ket{2x+1,2y} \right) \label{psi_1}
\end{eqnarray}
with $\omega=\exp({{\pi i}/{n}})$.  Variable $k,l$  run from 0 to $2n-1$.
For fixed values of $k$ and $l$, those vectors define a plane that is invariant under the action of $U$, that is
\begin{eqnarray}
U\ket{{\psi}_{k l}^{\textrm{0}}} & = &A_{k l}\ket{{\psi}_{k l}^{\textrm{0}}}- B_{k l}^*\ket{{\psi}_{k l}^{\textrm{1}}},\label{eq:Hpsitilde1}\\
U\ket{{\psi}_{k l}^{\textrm{1}}} & = &B_{k l}\ket{{\psi}_{k l}^{\textrm{0}}}+A_{k l}^*\ket{{\psi}_{k l}^{\textrm{1}}},\label{eq:Hpsitilde2}
\end{eqnarray}
where $A_{k l}=a_{k l}+i\,b_{k l}$,  $B_{k l}=c_{k l}+i\,d_{k l}$ , and
\begin{eqnarray}
a_{k l}&=&\frac{1}{2}\,\big(   \cos \tilde{k}+ \cos \tilde{l}\big)^2-1   ,\label{a_kl}\\ 
b_{k l}&=&-\frac{1}{2}\,\big(\sin \tilde{k}+\sin \tilde{l}\big)\big(\cos \tilde{k}+\cos \tilde{l}\big),\label{b_kl}\\
c_{k l}&=&\frac{1}{2}\,\sin(\tilde{l}-\tilde{k})\big(\cos \tilde{k}+\cos \tilde{l}\big),\\
d_{k l}&=&\frac{1}{2}\,\big(\cos(\tilde{k}-\tilde{l})-1\big)\big(\cos \tilde{k}+\cos \tilde{l}\big).
\end{eqnarray}
The new tilde variables  are $\tilde{k}={\pi k}/{n}$ and $\tilde{l}={\pi l}/{n}.$

The analysis of the dynamics can be reduced to a two-dimensional subspace of ${\cal H}^N$ by defining a reduced evolution operator 
\begin{equation}
U_{\textrm{RED}}^{(k l)}=\left[\begin{array}{cc}
 A_{k l} & B_{k l} \\
  - B_{k l}^* &  A_{k l}^*
\end{array}\right],
\end{equation}
which is unitary because $A_{k l}\, A_{k l}^{*}+B_{k l}\, B_{k l}^{*}=1.$ A vector in this two-dimensional subspace is mapped to the Hilbert space ${\cal H}^N$ after multiplying its first entry by $\ket{{\psi}_{k l}^{\textrm{0}}}$ and its second entry by $\ket{{\psi}_{k l}^{\textrm{1}}}$. 

Now we show that an eigenbasis of $U$ can be found from an eigenbasis of $U_{\textrm{RED}}^{(k l)}$. In fact, the eigenvalues of $U_{\textrm{RED}}^{(k l)}$ for $0\le k, l<2n$ are exactly the  eigenvalues of $U$, and if $ \ket{v_{k l}^\phi}$ is an eigenvector of  $U_{\textrm{RED}}^{(k l)}$ associated with eigenvalue  $\exp({i{\phi_{k l}}})$ then the corresponding eigenvector of $U$ is
\begin{equation} \label{psi_kl}
 \ket{\psi_{k l}^\phi}= \braket{0}{v_{k l}^\phi}\,\ket{\psi^{0}_{kl}} +     \braket{1}{v_{k l}^\phi}\,\ket{\psi^{1}_{kl}} , \\  
\end{equation}
where $\ket{\psi^{0}_{kl}}$ and $\ket{\psi^{1}_{kl}}$ are given by Eqs.~(\ref{psi_0}) and~(\ref{psi_1}).

The eigenvalues of $U_{\textrm{RED}}^{(k l)}$ are $\exp({i{\phi_{k l}}})$ for $0\le k<n $ and  $\exp({-i{\phi_{k l}}})$ for $n\le k<2n$, where 
\begin{equation}\label{theta_kl}
\phi_{k l} =	
 \begin{cases} \pi, & \mbox{if } k\pm l\equiv n \mod (2n),\\  
-2\pi{k}/n, & \mbox{if } k=l,  \\  
\arccos(a_{kl}), & \mbox{otherwise.} \end{cases}
\end{equation}
The corresponding normalized eigenvectors in the nontrivial cases are
\begin{equation}\label{v_theta_kl}
  \ket{v_{k l}^\phi}=\frac{1}{\sqrt{2\sin \phi_{k l} }}\left[ \begin {array}{c} {\sqrt { b_{k l}+  \sin \phi_{k l}  }}\\ \noalign{\medskip}{\frac { d_{k l}+ic_{k l} }{\sqrt { b_{k l}+ \sin \phi_{k l} }}}\end {array} \right]
\end{equation}
for $0\le k<n $ and  $\ket{v_{k l}^{-\phi}}$ for $n\le k<2n$. When $k=l$ or $k\pm l\equiv n \mod (2n)$, the corresponding eigenvectors are $\ket{0}=\left[ \begin {array}{c} {1}\\ {0}\end {array} \right]$, if $0\le k<n$ and $\ket{1}=\left[ \begin {array}{c} {0}\\ {1}\end {array} \right]$,  if $n\le k<2n$.  From the characterization  of these eigenvalues and eigenvectors of $U_{\textrm{RED}}^{(k l)}$,  we can obtain the eigenvalues and an orthonormal eigenbasis of $U$, which are described in Table~\ref{tab:basisU}.

\begin{table}[H]
\centering
\begin{tabular}{| c | c |p{4.3cm}| } 
  \hline			
  Eigenvalue & Eigenvector & $(k,l)\mod 2n$ \\
  \hline
  $-1$ & $\ket{\psi^{0}_{kl}}$& if  $0\le k<n$ and  $k\pm l \equiv n $ \\
  $-1$ & $\ket{\psi^{1}_{kl}}$&  if  $n\le k<2n$ and  $k\pm l \equiv n $ \\
  $\textrm{e}^{-\frac{2\pi{k}i}{n}}$& $\ket{\psi^{0}_{kl}}$ &  if  $0\le k<n$ and  $k=l$ \\
  $\textrm{e}^\frac{2\pi{k}i}{n}$ & $\ket{\psi^{1}_{kl}}$  &  if  $n\le k<2n$ and  $k=l $ \\
  $\textrm{e}^{\pm i \phi_{k l} }$ & $\ket{\psi_{k l}^{\pm\phi}}$ & otherwise\\
  \hline
\end{tabular}
\caption{Eigenvalues and eigenvectors of $U$, where  $\ket{\psi^{0}_{kl}}$, $\ket{\psi^{1}_{kl}}$, and   $\ket{\psi_{k l}^{\pm\phi}}$ are given by Eqs.~(\ref{psi_0}),~(\ref{psi_1}), and~(\ref{psi_kl}), respectively.}
\label{tab:basisU}
\end{table}


\section{Analysis of the algorithm}\label{sec3}

In order to determine the efficiency of our algorithm, we need to find the running time and the success probability. The optimal running time is the number of steps $t$ that corresponds to the first maximum of the success probability.  The success probability is $\left|\bracket{0,0}{({\cal U}_0)^t}{\psi_0}\right|^2$, where $\ket{\psi_0}$ is the initial state given by Eq.~(\ref{IC}) and $\ket{0,0}$ is the target or marked state.

To calculate $\left|\bracket{0,0}{({\cal U}_0)^t}{\psi_0}\right|^2$, we will write $\ket{\psi_0}$ and $\ket{0,0}$ in the eigenbasis of ${\cal U}_0$. Only two eigenvectors play a relevant role in this analysis. The same procedure is used to analyze the Grover algorithm, which also depends on only two eigenvectors of the modified evolution operator. The first one is the eigenvector associated with the eigenvalue with the smallest positive argument and the second one is its complex conjugate~\cite{Portugal:Book}. The state of the quantum computer running the Grover algorithm is an exact superposition of those two eigenvectors. In our algorithm, the state of the quantum walk will be approximately described by the superposition of the eigenvector of ${{\cal U}_0}$ associated with the eigenvalue with the smallest argument and a second eigenvector, which is not the complex conjugate the first one. Similar approaches were used in coined walks on lattices~\cite{HT09,Hein:2010}.

Let $\exp(i \lambda)$ be the eigenvalue of ${\cal U}_0$ with the smallest positive argument $\lambda$ and let $\ket{\lambda}$ be its associated eigenvector, that is, ${\cal U}_0\ket{\lambda}=\exp(i \lambda)\ket{\lambda}$. We now describe a method to calculate $\lambda$ using the spectrum of $U$. Recall that $U$ is the evolution operator with no marked elements.

 Let $\ket{\psi_{k l}^{\phi}}$ represent a generic eigenvector of $U$, as described in Table~\ref{tab:basisU}, associated with eigenvalue $\exp(i \phi_{k l})$, where $\phi_{k l}$ is given by Eq.~(\ref{theta_kl}) (the sign of $\phi_{k l}$ inverts if $n\le k<2n$). Using the completeness relation, we have
\begin{equation}\label{bra00lambda}
 \braket{0,0}{\lambda}\,=\,\sum_{kl} \braket{0,0}{\psi_{k l}^{\phi}}\braket{\psi_{k l}^{\phi}}{\lambda},
\end{equation}
where the sum runs over all values of $(k,l)$. On the other hand, from the expression $\bracket{\psi_{k l}^{\phi}}{{\cal U}_0}{\lambda}=\bracket{\psi_{k l}^{\phi}}{U R_0}{\lambda}$, we obtain
\begin{equation}\label{psi_lambda}
 \braket{\psi_{k l}^{\phi}}{\lambda}\,=\,\frac{2\braket{0,0}{\lambda}\braket{\psi_{k l}^{\phi}}{0,0}}{1-\textrm{e}^{i(\lambda-\phi_{k l})}}.
\end{equation}
Using the above equation in (\ref{bra00lambda}) and $\braket{0,0}{\psi_{k l}^{\phi}}=\braket{0}{v_{k l}^{\phi}}/\sqrt 2n$, we obtain
\begin{equation}
 \sum_{k l} \frac{\left|\braket{0}{v_{k l}^\phi}\right|^2}{1-\textrm{e}^{i(\lambda-\phi_{k l})}} =n^2,
\end{equation}
which is valid if $\lambda\neq \phi_{k l}$ for all $k,l$. Using that $2/(1-\textrm{e}^{ia})=1+i\sin a/(1-\cos a)$, the imaginary part of the above equation reads as
\begin{equation}\label{exactsum_kl}
 \sum_{k l} \left|\braket{0}{v_{k l}^\phi}\right|^2\frac{\sin({\lambda-\phi_{k l}})}{1-\cos(\lambda-\phi_{k l})} =0.
\end{equation} 
Using Eq.~(\ref{theta_kl}), the left hand side of~(\ref{exactsum_kl}) splits into three terms 
\begin{eqnarray}\label{sum3terms}
 (1-2n)\tan\frac{\lambda}{2}+
   \sum_{\substack{k=0 \\  2k \neq n }}^{n-1} \frac{\sin (\lambda+2\tilde{k})}{1-\cos (\lambda+2\tilde{k})} + & &  \nonumber \\
  \sum_{\substack{k,l=0  \\ k\pm l \not\equiv n,\, k \neq l }}^{2n-1} \left|\braket{0}{v_{k l}^\phi}\right|^2 \frac{\sin({\lambda-\phi_{k l}})}{1-\cos(\lambda-\phi_{k l})}&=&0.
\end{eqnarray}
This equation can be used to calculate $\lambda$ by means of numerical methods. In order to proceed analytically, we suppose that $\lambda\ll \phi_\textrm{min}$ for $n\gg 1$, where $\phi_\textrm{min}$ is the smallest positive value of $\phi_{k l}$. We will check the validity of this assumption later.

Assuming $\lambda\ll \phi_\textrm{min}$ for large $n$ and disregarding terms quadratic in $\lambda$, Eq.~(\ref{sum3terms}) reduces to
\begin{equation}\label{eq:22}
\frac{1}{\lambda}-n^2C^2\lambda=O\left(\lambda^2\right)
\end{equation}
where
\begin{eqnarray}\label{B2}
 C^2 &=& \frac{1}{2n^2}\sum_{\substack{k,l=0  \\ k\pm l \not\equiv n,\, k \neq l }}^{2n-1} \frac{\left|\braket{0}{v_{k l}^\phi}\right|^2}{1-\cos\phi_{k l}} + O(1).
\end{eqnarray}
Up to first order in $\lambda$, the solutions of Eq.~(\ref{eq:22}) are
\begin{equation}\label{lambda_main}
\lambda\,=\,\pm\frac{1}{nC}.
\end{equation}
Those solutions show that both $\exp(\pm i \lambda)$ are the eigenvalues of ${\cal U}_0$. In the Appendix we show that $C=\Theta(\sqrt{\ln n})$. Therefore, $1/\lambda=\Theta(\sqrt{N\ln N})$. Using Eq.~(\ref{theta_kl}), we verify that $\phi_\textrm{min}$ is attained when $(k,l)=(1,0)$, which shows that $\phi_\textrm{min}=\Theta(1/\sqrt{N})$, confirming  that $\lambda\ll \phi_\textrm{min}$ for $n\gg 1$ is a valid approximation.

Now writing the target state $\ket{0,0}$ in the eigenbasis of ${\cal U}_0$, we obtain
\begin{equation}\label{ket00}
	\ket{0,0}= \braket{\lambda}{0,0}\,\ket{\lambda}+ \braket{\lambda^-}{0,0}\,\ket{\lambda^-}+\ket{\lambda^\perp},
\end{equation}
where $\ket{\lambda^-}$ is the eigenvector of ${\cal U}_0$ associated with $\exp(- i \lambda)$ and $\ket{\lambda^\perp}$ is the component of $\ket{0,0}$ orthogonal to the plane spanned by $\ket{\lambda}$ and $\ket{\lambda^-}$. We do not need to know the full expressions of $\ket{\lambda}$ or  $\ket{\lambda^-}$ in this analysis. Using  Eq.~(\ref{psi_lambda}) in the normalization condition $\sum_{k l}  \left|\braket{\psi_{k l}^{\phi}}{\lambda}\right|^2=1$ and $\left|\braket{\psi_{k l}^{\phi}}{0,0}\right|^2=\left|\braket{0}{v_{k l}^{\phi}}\right|^2/2n^2$, we obtain
\begin{equation}
	\frac{1}{\left|\braket{0,0}{\lambda}\right|^2} = \frac{1}{n^2} \sum_{k l}  \frac{\left|\braket{0,0}{v_{k l}^{\phi}}\right|^2}{1-\cos\left(\lambda-\phi_{k l}\right)}.
\end{equation}
Expanding the sum into three terms similar to what we have done in Eq.~(\ref{sum3terms}), assuming that $\lambda\ll\phi_\textrm{min}$ for $n\gg 1$,  and keeping the dominant terms, we obtain
\begin{eqnarray}\label{ket00lambda_v2}
	\frac{1}{\left|\braket{0,0}{\lambda}\right|^2}&=&\frac{2}{n^2\lambda^2}+ \frac{1}{n^2}\sum_{\substack{k,l=0  \\ k\pm l \not\equiv n,\, k \neq l }}^{2n-1}  \frac{\left|\braket{0,0}{v_{k l}^{\phi}}\right|^2}{1-\cos \phi_{k l} }\nonumber \\
	&+&   \,O\left(1\right).
\end{eqnarray}
Using Eqs.~(\ref{lambda_main}) and~(\ref{B2}), we obtain
\begin{equation}\label{ket00lambda}
	\frac{1}{\left|\braket{0,0}{\lambda}\right|^2} =\frac{4}{n^2\lambda^2}+O\left(1\right).
\end{equation}
Without loss of generality, we assume that $\braket{0,0}{\lambda}$ is a positive real number. In fact, if $\braket{0,0}{\lambda}=a\,\textrm{e}^{ib}$, where $a$ and $b$ are real numbers and $a$ is positive, we redefine $\ket{\lambda}$ as  $\textrm{e}^{-ib}\ket{\lambda}$. After this redefinition,$\braket{0,0}{\lambda}$ is a positive real number given by $n\lambda/2+O(1)$. The same reasoning applies to $\braket{0,0}{\lambda^-}$, and we also obtain $\braket{0,0}{\lambda^-}=n\lambda/2+O(1)$.

Decomposing $\ket{\psi_0}$ in the eigenbasis of ${\cal U}_0$, we obtain
\begin{equation}\label{psi_0_lambda}
	\ket{\psi_0}= \braket{\lambda}{\psi_0}\,\ket{\lambda}+ \braket{\lambda^-}{\psi_0}\,\ket{\lambda^-}+\ket{\psi_0^\perp},
\end{equation}
where $\ket{\psi_0^\perp}$ is the component of $\ket{\psi_0}$ orthogonal to the plane spanned by $\ket{\lambda}$ and $\ket{\lambda^-}$. Using Eq.~(\ref{psi_0}), we verify that $\ket{\psi_0}=\ket{\psi^{0}_{kl}}$ for $(k,l)=(0,0)$. Using Eq.~(\ref{psi_lambda}) with $(k,l)=(0,0)$, we obtain
\begin{equation}\label{lambda_psi}
	\braket{\lambda}{\psi_0}=-\frac{i\,\textrm{e}^{\frac{i\lambda}{2}}}{\sqrt 2} +O\left(\lambda^2\right).
\end{equation} 
Using Eq.~(\ref{psi_lambda}) with $(k,l)=(0,0)$ again, but this time replacing $\ket{\lambda}$ by  $\ket{\lambda^-}$, we obtain that $\braket{\lambda^-}{\psi_0}=\left(\braket{\lambda}{\psi_0}\right)^*$. It is straightforward to check that
\begin{equation}
\left|\braket{\lambda}{\psi_0}\right|^2+\left|\braket{\lambda^-}{\psi_0}\right|^2=1+O\left(\lambda^2\right).
\end{equation}
Then, we can ignore the term $\ket{\psi_0^\perp}$ in Eq.~(\ref{psi_0_lambda}).

Using (\ref{psi_0_lambda}) and (\ref{lambda_psi}), we obtain
\begin{eqnarray}
	{({\cal U}_0)^t}\ket{\psi_0} &=&  \left(-\frac{i\textrm{e}^{i\lambda\left(t+\frac{1}{2}\right)}}{\sqrt 2}+O\left(\lambda^2\right)\right)\ket{\lambda}+\nonumber \\
	&&\left(\frac{i\textrm{e}^{-i\lambda\left(t+\frac{1}{2}\right)}}{\sqrt 2} +O\left(\lambda^2\right)\right)\ket{\lambda^-}.
\end{eqnarray}
Using (\ref{ket00}) and (\ref{ket00lambda}), we obtain
\begin{equation}\label{ket00Upsi0}
	\left|\bracket{0,0}{({\cal U}_0)^t}{\psi_0}\right|^2=
	\frac{n^2\lambda^2}{2}\, \sin^2 \lambda\left(t+\frac{1}{2}\right)+O\left(\lambda^2\right).
\end{equation}
The success probability is
\begin{equation}
	P = \frac{n^2\lambda^2}{2}+O\left(\lambda^2\right)
\end{equation}
and the running time is the first value of $t$ that maximizes the right hand side of Eq.~(\ref{ket00Upsi0}) ignoring terms $O(\lambda^2)$, which is
\begin{equation}
	t\,=\, \frac{\pi}{2\lambda}.
\end{equation}
Since $1/\lambda=\Theta(\sqrt{N\ln N})$, the success probability is $P=\Theta(1/\ln N)$ and the running time is $t=\Theta(\sqrt{N\ln N})$.

There are three possible ways to improve the success probability: (1) If we use the amplitude-amplification method, the time complexity of the algorithm is the original running time times $\sqrt{P}$, which yields $\Theta\big(\sqrt{N}\ln N\big)$ with success probability $O(1)$~\cite{BBHT98,Portugal:Book}. (2) If we add an extra qubit to the system, we can use Tulsi's method~\cite{Tul12} and the time complexity of the algorithm would be $\Theta(\sqrt{N\ln N})$. (3) We can use the results of Ref.~\cite{Ambainis:2012}, which showed that after running the quantum algorithm, the walker is close enough to the marked location so that a classical post-processing search using the overhead time $O\big(\sqrt{N}\big)$ is enough to find the marked vertex with probability $O(1)$. This means that we do not need to use amplitude amplification and the time complexity of the algorithm is $\Theta(\sqrt{N\ln N})$ with success probability $O(1)$.

\section{Numerical analysis of alternative models}

The evolution operator of the staggered quantum walk model with Hamiltonians~\cite{POM16} is the product of local operators $U_{ab} = \textrm{e}^{i\,\theta H_{ab}}$, where $\theta$ is an angle. In the previous sections, we have analyzed the case $\theta=\pi/4$ and $U =- U_{11} U_{10} U_{01} U_{00}$, where $H_{a b}$ is given by Eq.~(\ref{eq:H}). Since we have obtained analytical expressions for the running time and success probability, we will not show numerical results for this case. On the other hand, we consider alternative quantum walks on the two-dimensional lattice based on the staggered model that can be obtained by permuting the order of the tessellations, by changing the value of $\theta$, or by choosing non-uniform polygons. For those alternative cases, we have performed numerical calculations employing program Hiperwalk~\cite{LLP15} using high performance computing on Nvidia Tesla cards K20 and K40.

\begin{figure}[!h]
\centering
\includegraphics[trim=30 0 0 0,clip,scale=0.42]{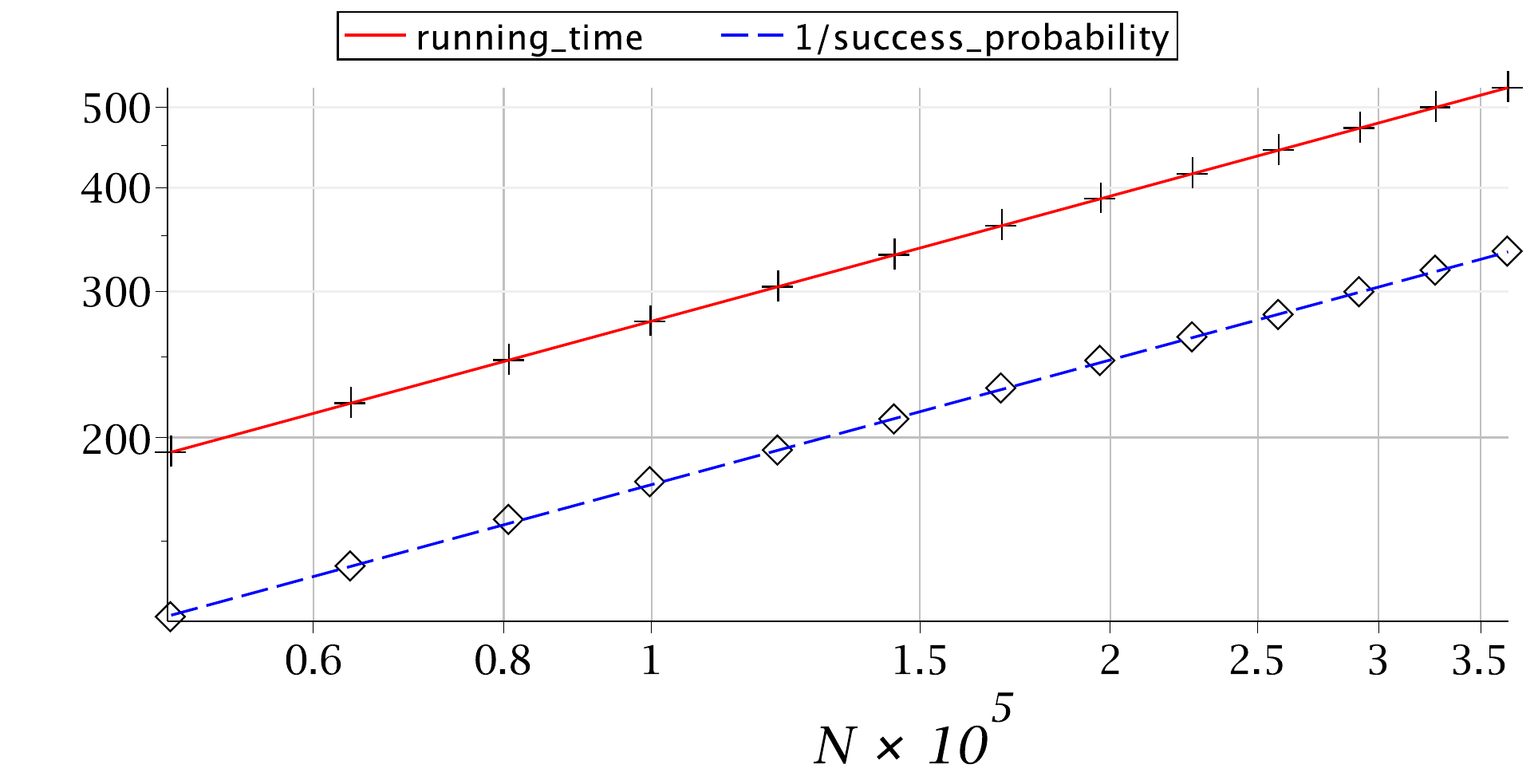}
\caption{(Color online) Running time (crosses, red line) and the inverse of the success probability (diamonds, blue dashed-line) as a function of $N$ in loglog scale when $\theta=\pi/4$ for the alternative model. The points are obtained from numerical simulations and the fitting lines using the least square method.}
\label{fig:TxN_0}
\end{figure}

Consider in this paragraph the alternative staggered model $U =-U_{11} U_{01} U_{10} U_{00}$, which inverts the order of the local operators keeping the same value for $\theta$, that is, $\theta=\pi/4$. Fig.~\ref{fig:TxN_0} depicts the running time and the inverse of the success probability as a function of the number of vertices $N$ in loglog scale. The analytical formula for the fitting lines are $276N^{0.500}$ for the running time and  $175N^{0.499}$ for the inverse probability, approximately. Those results suggest that the running time is $O\big(\sqrt{N}\big)$ and the success probability is $O(1/\sqrt{N})$. The success probability falls down too fast and we cannot use the results of Ref.~\cite{Ambainis:2012}. On the other hand, we can use the amplitude-amplification method in order to obtain a quantum algorithm with time complexity $O(N^{3/4})$ and success probability $O(1)$. This result is interesting because it is the only model, as far as we know, for the two-dimensional lattice, whose running time is $O\big(\sqrt{N}\big)$. In all other models, the running time is $O(\sqrt{N\ln N})$.

We have numerically analyzed values of $\theta$ different from $\pi/4$. As soon as we move away from  $\theta=\pi/4$, the time complexity becomes worse and approaches to $O(N\ln N)$. This is especially valid when $\theta=\pi/2$, which characterizes the standard staggered model~\cite{PSFG15}. We have also analyzed the dynamics of models with non-uniform vectors, that is, the vectors associated with the polygons have non-uniform amplitudes, in contrast to the vectors given by Eq.~(\ref{eq:u}) which have uniform amplitudes. When $\theta=\pi/2$, the algorithm speed is as slow as the speed of random-walk-based algorithms for any choice of amplitudes.

\begin{figure}[!h]
\centering
\includegraphics[trim=30 420 100 50,clip,scale=0.4]{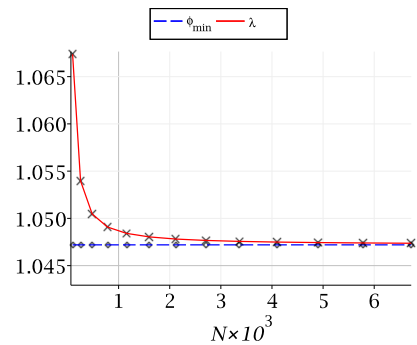}
\caption{(Color online) $\lambda$ (crosses, red line) and $\phi_{\textrm{min}}$ (diamonds, blue dashed-line) as a function of $N$ when $\theta=\pi/3$. The points are obtained from numerical calculations of the eigenvalues of $\mathcal{U}_0$ and $U$, respectively.}
\label{fig:lambda_phi_N}
\end{figure}

To understand why quantum walk searching has a bad behavior when $\theta\neq \pi/4$, we analyze the behavior of the eigenvalues of $U$ and ${\mathcal{U}}_0$ with the smallest positive arguments \big($\exp(i\phi_{\textrm{min}})$ and $\exp(i\lambda)$\big). Fig.~\ref{fig:lambda_phi_N} shows $\phi_{\textrm{min}}$ and $\lambda$ as a function of $N$ when $\theta=\pi/3$. Notice that both parameters tend to a constant value. The same result is valid for any other value of $\theta$ and the limiting constant is $|4\theta-\pi|$. Therefore, $\phi_{\textrm{min}}=\lambda=\Theta(1)$ when $\theta\neq \pi/4$. This is in stark contrast to the case $\theta=\pi/4$. In fact, in Section~\ref{sec3}, we have shown analytically that $\phi_{\textrm{min}}=\Theta(1/\sqrt{N})$ and $\lambda=\Theta(1/\sqrt{N\ln N})$, when $\theta=\pi/4$. We were able to employ the methods of Section~\ref{sec3} to calculate the running time and success probability because $\lambda\ll \phi_{\textrm{min}}$ asymptotically. Since the behavior of the eigenvalues with the smallest arguments plays a central role in determining the running time and success probability, the fact that $\phi_{\textrm{min}}$ and $\lambda$ are equal asymptotically shows that $U$ and ${\mathcal{U}}_0$ have the same ability to find the marked vertex when $\theta\neq\pi/4$. Operator $U$ cannot find the marked vertex; ${\mathcal{U}}_0$ cannot either.

\section{Conclusions}\label{conc}

We have described a new search algorithm in the two-dimensional lattice with $N$ vertices and cyclic boundary conditions in time $O(\sqrt{N\ln N})$ with success probability $O(1)$ using a staggered quantum walk with Hamiltonians. We have analytically proved that after $\Theta(\sqrt{N\ln N})$ time steps, the marked element is found with probability $\Theta(1/\ln N)$. Using the results of Ref.~\cite{Ambainis:2012}, a classical post-processing search with time $O\big(\sqrt{N}\big)$ is enough to find the marked vertex with success probability $O(1)$.

We highlight that it is possible to reach the results of the present paper only because we have used the staggered model with Hamiltonian with $\theta=\pi/4$. The staggered model with Hamiltonians generalizes the standard staggered model, is more amenable for experimental implementations~\cite{coinless}, and has other interesting features such as perfect state transfer~\cite{CP17}. On the other hand, numerical implementations show that the time complexity of search algorithms based on the standard staggered model ($\theta=\pi/2$) are as bad as random-walk-based algorithms.


\section*{Acknowledgements}
The authors acknowledge financial support from Faperj and CNPq.

\

\section*{Appendix}

In this Appendix we show that $C$ (Eq.~(\ref{B2})) is $\Theta(\sqrt{\ln n})$. We can show that 
\begin{equation}\label{B10}
  \sum_{\substack{k,l=0  \\ k\pm l \not\equiv n,\, k \neq l }}^{2n-1} \frac{b_{k l}}{\sin\phi_{k l}\,(1-\cos\phi_{k l})} \,=\,0
\end{equation}
by using the symmetry $g({k, l})\equiv-g({n+k,n+l})\mod 2n$, for $0\le k<2 n$ and $0\le l < n$, where $g(k,l)$ is the summand of the sum on the left hand side of the above equation.

Using Eq.~(\ref{B10}) and the first entry of $\ket{v_{k l}^\phi}$ (Eq.~(\ref{v_theta_kl})), the expression of $C^2$ (Eq.~(\ref{B2})) reduces to
\begin{equation}\label{B11}
 C^2 \,=\, \frac{1}{4n^2}\sum_{\substack{k,l=0  \\ k\pm l \not\equiv n,\, k \neq l }}^{2n-1} \frac{1}{1-\cos\phi_{k l}} + O(1).
\end{equation}
Since the sum of terms obeying $k\pm l\equiv n\mod 2n$ and $k=l$ (when $(k,l)\neq(0,0)$ and $(k,l)\neq(n,n)$) over $n^2$ are $O(1)$, we can add those terms to the sum. Using Eq.~(\ref{theta_kl}), we obtain
\begin{equation}\label{B11a}
 C^2 \,=\,  \frac{1}{2n^2}\sum_{\substack{k,l=0  \\ (k,l)\ne (0,0) \\ (k,l) \neq (n,n) }}^{2n-1} \frac{1}{4-f_{k l}^2} + O(1),
\end{equation}
where
\begin{equation}\label{B3a2}
 f_{k l}=\cos \frac{\pi k}{n}+\cos \frac{\pi l}{n}.
\end{equation}
When the summand is $1/(4-f_{k l}^2)$, we can split the sum into four terms
\begin{equation}\label{B3b}
 \sum_{\substack{k,l=0  \\ (k,l)\ne (0,0)  }}^{n-1} +
 \sum_{k=0,l=n}^{n-1,2n-1} +
 \sum_{k=n,l=0}^{2n-1,n-1} +
 \sum_{\substack{k,l=n  \\ (k,l) \neq (n,n) }}^{2n-1} 
\end{equation}
and analyzing the range and relabeling the dummy indices, we conclude that~(\ref{B3b}) is equal to
\begin{equation}\label{B3c}
 \frac{1}{2}+4\sum_{\substack{k,l=0  \\ (k,l)\ne (0,0)  }}^{n-1}
 \frac{1}{4-f_{k l}^2}. 
\end{equation}
Then,
\begin{equation}\label{B12}
 C^2 \,=\,  \frac{2}{n^2}\sum_{\substack{k,l=0  \\ (k,l)\ne (0,0)  }}^{n-1} \frac{1}{4-f_{k l}^2} + O(1).
\end{equation}
Using that $4-f_{k l}^2=(2-f_{k l})(2+f_{k l})$ and the fact that
\begin{equation}\label{B3e}
 \frac{2n^2}{3}-\frac{n}{\sqrt 2}+
 \sum_{\substack{k,l=0  \\ k,l\ne 0,0 }}^{n-1} \frac{1}{2+f_{k l}} 
  =
 \sum_{\substack{k,l=0  \\ k,l\ne 0,0 }}^{n-1} \frac{1}{2-f_{k l}} + O(1), 
\end{equation}
we obtain
\begin{equation}\label{B13}
 C^2 \,=\,  \frac{1}{n^2}\sum_{\substack{k,l=0  \\ (k,l)\ne (0,0)  }}^{n-1} \frac{1}{2-f_{k l}} + O(1).
\end{equation}
Using that 
\begin{equation}
1-\frac{\pi^2 k^2}{2\,n^2} \le \cos\left(\frac{\pi k}{n}\right)\le 1-\frac{2k^2}{n^2}
\end{equation}
for $0\le k<n$, we obtain, up to order $O(1)$,
\begin{eqnarray}\label{B6}
\frac{2\, I(n)}{\pi^2}\,\le\, C^2 \,\le\, \frac{I(n)}{2},
\end{eqnarray}
where 
\begin{equation}\label{I_n}
 I(n)\\=\,\sum_{\substack{k,l=0  \\ (k,l)\neq (0,0) }}^{n-1} \frac{1}{k^2+l^2},
\end{equation}
Notice that we have to add constant terms to~(\ref{B6}) in order to obtain valid inequalities for small $n$. The sum on the right hand side of Eq.~(\ref{I_n}) has been addressed in~Ref.~\cite{Ambainis:2005}, which proved that it is $\Theta(\sqrt{\ln n})$. This shows that $C=\Theta(\sqrt{\ln n})$.


\end{document}